\documentclass[sn-nature]{sn-jnl}
\usepackage{graphicx}%
\usepackage{amsmath,amssymb,amsfonts}%
\usepackage{amsthm}%
\usepackage{mathrsfs}%
\usepackage[title]{appendix}%
\usepackage{xcolor}%
\usepackage{textcomp}%
\usepackage{manyfoot}%
\usepackage{booktabs}%
\usepackage{algorithm}%
\usepackage{algorithmicx}%
\usepackage{algpseudocode}%
\usepackage{listings}%

\theoremstyle{thmstyleone}

\theoremstyle{thmstyletwo}%
\theoremstyle{thmstylethree}%
\usepackage{longtable}

\begin{document}
\title[Science Growth]{Growth of Science: How long will the United States uphold its position?}

\author*[1]{\fnm{Dipak} \sur{Patra}}\email{dipak@rri.res.in}

\affil*[1]{\orgdiv{Soft Condensed Matter Group}, \orgname{Raman Research Institute}, \orgaddress{\street{C. V. Raman Avenue, Sadashivanagar}, \city{Bangalore}, \postcode{560080}, \state{Karnataka}, \country{India}}}
\date{\today}

\abstract{
Policymakers often assess the growth of science in a country and compare it with that of other countries to set future planning for scientific research focusing on the sustainable development and economic growth of the country.
 Here, we study the growth of science for the period of 1996-2020 corresponding to the top fifty countries with the highest publications in 2020. It is found that the annual growth rates of scientific and technical journal publications exhibit Taylor's power law behavior indicating the dependence of the variance on the mean growth rate and the distributions of annual growth rates follow skew-symmetric distributions. Furthermore, we have computed the entropy based on annual publication numbers among the countries to assess the spatial disparity in the system. The entropy is found to increase mostly linear with time reducing the disparity among the countries. By performing the linear regression analysis, we predict that around the year 2046, all the countries excluding China may equally contribute towards the growth of science.
 We have also assessed the stability of the current USA ranking by computing the entropy between the USA and other countries. Based on the regression analysis, it is estimated that three potential countries such as Indonesia, India, and Iran may take the ranks ahead of the USA around the years 2024, 2029, and 2041 respectively.
}

\maketitle

\section{Introduction}
Scientific research is an essential part of a country impacting various fields including sustainable development, economic growth, and accumulating knowledge \citep{Matia_2005, Gupta_2013}. Indeed, the outcome of the research leads to technological advancement and provides the country with an upper hand towards leadership in the world. Therefore, countries across the globe sanction funds in the economic budget focusing on scientific research and development.
To allocate the funds, it is crucial to know the current trend of growth of science for a country. Nevertheless, the growth of science in the country needs to be critically assessed and compared with that of other countries to set future planning \citep{NSB_2020}. Therefore, the study of the growth of science gains worldwide attention after the pioneering investigation by the theoretician of science, Derek John de Solla Price \citep{Matia_2005, Bornmann_2021}. Since then a large number of studies have been carried out to investigate the growth of scientific research across the world \citep{Gupta_2013, NSB_2020, Asatani_2023, Coccia_2018, Patra_2006, Barth_2014, Pautasso_2012, Bornmann_2015, Javed_2018, Bornmann_2021, Matia_2005, Plerou_1999, Singh_2020}.

Most of these studies have utilized bibliometric data and measured the growth of science. The utilization of the bibliographic databases is advantageous compared to other existing data such as the number of researchers, amount of money allocated for research and development, number of research labs, etc., as these large-scale, multi-disciplinary databases are publicly available \citep{Bornmann_2021, NSB_2020}. Nevertheless, these databases are considered to be authentic as the scientific articles are mostly published after the peer-review process. Furthermore, scientists are always eager to publish their research work to improve their career trajectory and share their knowledge with others. Therefore, the bibliographic databases based on publication data can directly assess the growth of science in a country.
 
The growth of science in a country has mostly been studied by counting the number of publications on a yearly basis. It has been found that the number of publications grows either linearly or exponentially for most of the countries \citep{Bornmann_2015, Javed_2018, Bornmann_2021}. Therefore, regression analyses are often performed to find out the average growth rates of the countries and these growth rates are compared to each other to understand the global trend in the long-term \citep{NSB_2020}. Sometimes the annual growth rate is determined by comparing the number of publication data for two consecutive years. It should be noted that the annual growth rate fluctuates largely over time. An understanding of the fluctuation is necessary to stabilize the growth rate of science to a desired value. Only few studies deploying physics-based tools have assessed the fluctuation of annual growth rates for a given publication number by computing the conditional probability distribution and found that the standard deviation of the corresponding conditional distribution decreases with the publication number \citep{Matia_2005, Plerou_1999}. However, these studies do not describe how the fluctuation of growth rates depends on the mean growth rate for a country. Furthermore, it is well known that physics-based tools have been employed to study a wide variety of social phenomena such as the growth of business farm \citep{Amaral_1998, Lee_1998}, stock market \citep{Gopikrishnan_1998, Pan_2007}, crowd dynamics \citep{Helbing_2000, Silverberg_2013}, disease spreading \citep{Forgacs_2023}, sports \citep{Stock_2022, Patra_2024_cricket}, and determining population as well as crime center of a country \citep{Patra_2024_meancenter}. Surprisingly, despite the profound implications across various fields, physics-based tools have been less explored in the assessment of the growth of science.
Nevertheless, it has been observed that the number of year-wise publications widely varies across the countries leading to the spatial disparity \citep{NSB_2020, Gupta_2013, Singh_2020}. It is therefore worthwhile to ask how this disparity varies with time among the countries and when it will go away giving rise to equality in the publication output among these nations. However, these questions have not been addressed in the earlier studies as it is difficult to define a quantity that can measure the disparity. It should be noted that China suppressed the United States of America (USA) in the number of science publications in 2016 and continues to hold the top position. It is therefore interesting to ask how long the USA will uphold its second position in the world ranking. 

To address the issues discussed above, here, we study the growth of science in the top fifty countries using physics-based methods. For the assessment of the fluctuation of the growth rates, three types of annual growth rates are estimated from the number of year-wise publications including scientific and technical journal articles. The variances of the annual growth rates are found to depend on the mean growth rates. A quantity like Shanon entropy is constructed based on the number of year-wise publications to assess the disparity among the countries. Most of the time the entropy is found to increase linearly with time. The linear regression analyses are performed to determine the time when the countries may contribute equally to the year-wise total scientific publications. By computing entropy, we also perform comparative studies between the USA and other countries and assess the stability of the current rank of the USA.

\begin{table}\label{tab:countryrank}
\caption{List of the top fifty countries selected in this study based on year-wise publications including scientific and technical journal articles in 2020. }
\begin{tabular}{p{1.5in} p{1.8in} p{1.5in} }
Rank & Country & Publications  \\
\hline
1 & China & 669744.3 \\ 2 & USA &  455855.57\\
3 & India &  149212.62  \\ 4 & Germany  & 109378.75\\
5 & United Kingdom  & 105564.47 \\ 6 &Japan  &  101014.27\\
7& Russian Federation & 89967.04\\ 8 & Italy  & 85419.29\\
9 &Korea Rep & 72490.44 \\ 10 & Brazil & 70291.66 \\
11 &  France  & 66478.55\\ 12 &  Canada & 65821.57 \\
13  & Spain & 65638.38\\ 14 & Australia  & 60890.77 \\
15 &Iran & 57755.27\\  16 & Turkey &  42623.31 \\
17 & Poland & 36766.63 \\ 18 & Netherlands  & 33376.81\\
19 & Indonesia  & 32553.79\\ 20 & Switzerland & 23075.04 \\
21 & Malaysia & 21884.84\\ 22 & Sweden & 21880.49\\
23 & Mexico & 20074.24\\ 24 & Egypt & 18469.18\\
25 & Saudi Arabia & 17321.16\\ 26& Portugal & 17098.9\\
27 & Pakistan & 17038.18\\ 28 & Belgium & 17015.55\\
29 & South Africa & 15884.53\\ 30 & Denmark & 15318.52\\
31& Czechia & 14994\\ 32 & Thailand & 13963.09\\
33 & Israel & 13955.4\\ 34 & Austria & 13700.08\\
35 & Norway & 13262.07\\ 36 & Ukraine & 12776.85\\
37 & Greece & 12538.66\\ 38 & Singapore & 12221.19\\
39 & Finland & 11328.41\\ 40 & Iraq & 9814.39\\
41 & Argentina & 9729.75\\ 42 & Romania & 9710.46\\
43 & Colombia & 9296.64\\ 44 & New Zealand & 8983.48\\
45 & Chile & 8979.79\\ 46 & Ireland & 8562.78\\
47 & Vietnam & 8213.16\\ 48 & Nigeria & 7899.95\\
49 & Hungary & 7478.17\\ 50 & Morocco & 6452.54\\
\hline
\end{tabular}
\end{table}

\section{Method}
We have collected the year-wise publication data of countries for the years 1996 to 2020 from the World Bank data repository accessed on December 2023 \citep{World_Bank_2024}.
As can be seen in table~\ref{tab:countryrank}, fifty countries with the highest publications including scientific and technical journal articles in 2020 have been considered in this study. It should be noted that the year-wise publication numbers are positive real numbers according to the description provided by the World Bank. We have computed three types of annual growth rates corresponding to each country which are described below. The relative annual growth of a country is defined as $
r_g = \frac{N(t)}{N(t-1)}$
 where $N(t)$ is the number of publications at time $t$.
 The actual and the logarithmic growth rates are related to the relative growth rate $r_g$ as
$r_a = r_g -1$, and $r_l = ln r_g$,
 respectively. The growth rate $r_g$ is always a positive number whereas the other two growth rates $r_a$ and $r_l$ are real numbers. The average growth rates and their variances of a country are calculated as 
\begin{eqnarray*}
\mu_x  &=&\sum_{i=1}^{T} \frac{r_{ix}}{T} ~,
\end{eqnarray*}
and
\begin{eqnarray*}
 \sigma^2_x &=& \sum_{i=1}^{T} \frac{(r_{ix} -\mu_x )^2}{T}~,
 \end{eqnarray*}
respectively, where $x \in \{g,a,l \}$, $T=2020-1996=24$ and $i$ counts the number of years.
To assess either the disparity among the countries we have calculated the entropy defined as
 \begin{eqnarray*}
S^\prime(t)&=&-\sum_k^{M} P_k(t)lnP_k(t)~,
\end{eqnarray*}
where $M$ is the number of countries which are included in the calculation and 
$P_k(t)=\frac{N_k(t)}{\sum_j^{M} N_j(t)}$ represents the fractional share of $k-$th country in the total publications of $M$ number of countries. In physics, this form of entropy expression is named Gibbs entropy whereas it is called Shanon entropy in statistics, machine learning, information theory, and image analysis.
One can also define a relative entropy as
 \begin{eqnarray*}
S(t)&=&-\frac{1}{lnM} S^\prime(t)~.
\end{eqnarray*}
It should be noted that the relative entropy $S$ takes values in a range between 0 and 1. For convenience, henceforth, we consider $S$ the entropy.
The maximum value of $S$ is equal to one which implies that all the countries possess the same number of publications giving rise to equality among these countries. In contrast, the minimum value of entropy equal to zero destroys equality and produces large disparity, giving rise to the monopoly in the system where only one country produces the total number of publications.
For convenience, we define two-time scales $t_{disp}$ and $t_{equal}$ as at time $t= t_{disp}$ the entropy equals to zero whereas it is equal to one for the time $t= t_{equal}$. 
We have performed regression analyses to describe the variation of entropy as a function of time.
To assess the quality of the fitting based on these analyses, we have computed the coefficient of determination, $R^2$, described as 
 \begin{eqnarray*}
R^2 &=& 1 -\frac{\sum_i^n (y_i -\hat{y}_i)^2}{\sum_i^n (y_i -\bar{y})^2} ~,
\end{eqnarray*}
where  $\hat{y}$ and $\bar{y}$ represent the predicted and mean of a variable $y$. The value of this statistical parameter close to one indicates a better fit.

\section{Result and Discussion}
\begin{figure}
  \includegraphics[clip=true,width=\columnwidth]{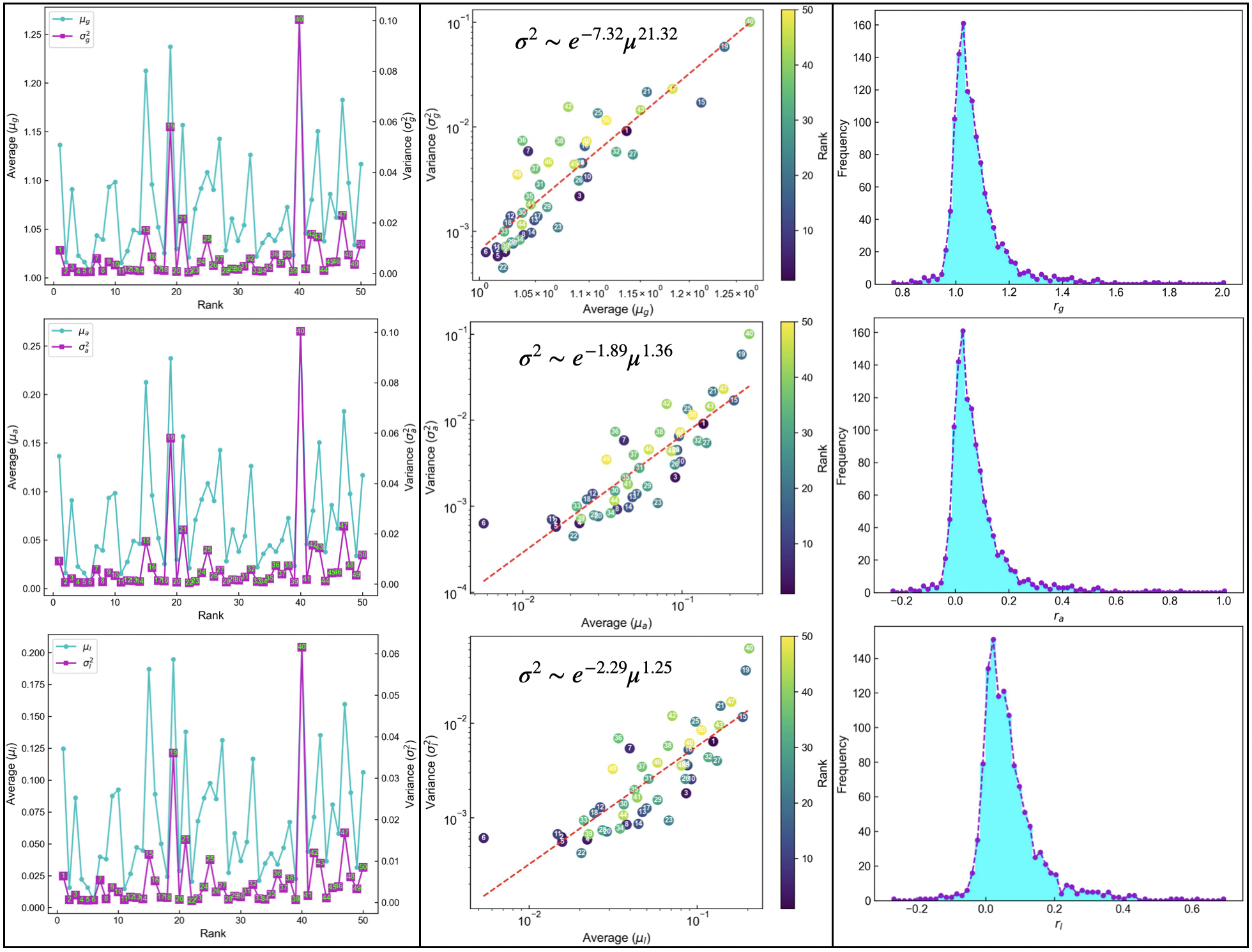}
   \caption{Top, middle, and bottom sections of each pan presenting the statistics of relative, actual, and logarithmic annual growth rates, respectively. The left pan indicating the variation of mean growth rates and respective variances with the current rank of the countries. Middle pan displaying the power-law dependence of the variance on the mean growth rate where both axes are in logarithmic scale. The dashed straight lines are produced based on the results from regression analyses. The equations corresponding to these straight lines are annotated. The right pan presenting the histograms corresponding to these three growth rates. }
    \label{fig:growth_rates}
\end{figure}
We have assessed the year-wise scientific and technical journal article publication data of the top fifty countries for the interval 1996-2020 by deploying various physics-based methods discussed above. The left pan of figure~\ref{fig:growth_rates} represents the variation of the mean annual growth rate and its variance as a function of the rank of the countries. It is found that both these statistical quantities randomly vary with the rank. For some countries especially Indonesia and Iraq, both these quantities are significantly large compared to other countries. It can be easily verified in the middle pan of figure~\ref{fig:growth_rates} which displays the variation of the variance with the mean annual growth rate. Interestingly, it is found that the variance $\sigma^2$ follows a power-law dependence on the mean growth rate $\mu$.
 The power-law dependence suggests that the countries with higher average growth rates show large variance giving rise to the large fluctuation over the publication output. Therefore, the growth rates of these countries are not stable over the years. 
 The power-law relation between the mean and variance described as $\sigma^2 \sim \mu^{\beta}$ is known as Taylor's power law where $\beta$ is the exponent. Taylor's power law has been empirically observed in various fields including biology, ecology, stock market, and physics (see the article \citep{Zoltan_2008} and references therein). To the best of our knowledge, for the first time, we have observed Taylor's power law in the growth of science. For most of the earlier observation, the exponent $\beta$ is nearly equal to $2$ such as $\beta=1.69$ for protein sample \citep{Vallania_2014}, $\beta=1.51$
for human chromosome 7 \citep{Kendal_2004}, and  $\beta=2$ for Wenchuan earthquake \citep{Shi_2019}.  However, any real value of $\beta$ is theoretically possible \citep{Giometto_2015}.
  Based on the regression analysis we have found that the power-law exponent $\beta$ corresponding to the growth of science takes the values $21.32, 1.36$, and $1.25$ for the relative, actual, and logarithmic growth rates respectively. 
  \begin{figure}[H]
  \includegraphics[clip=true,width=\columnwidth]{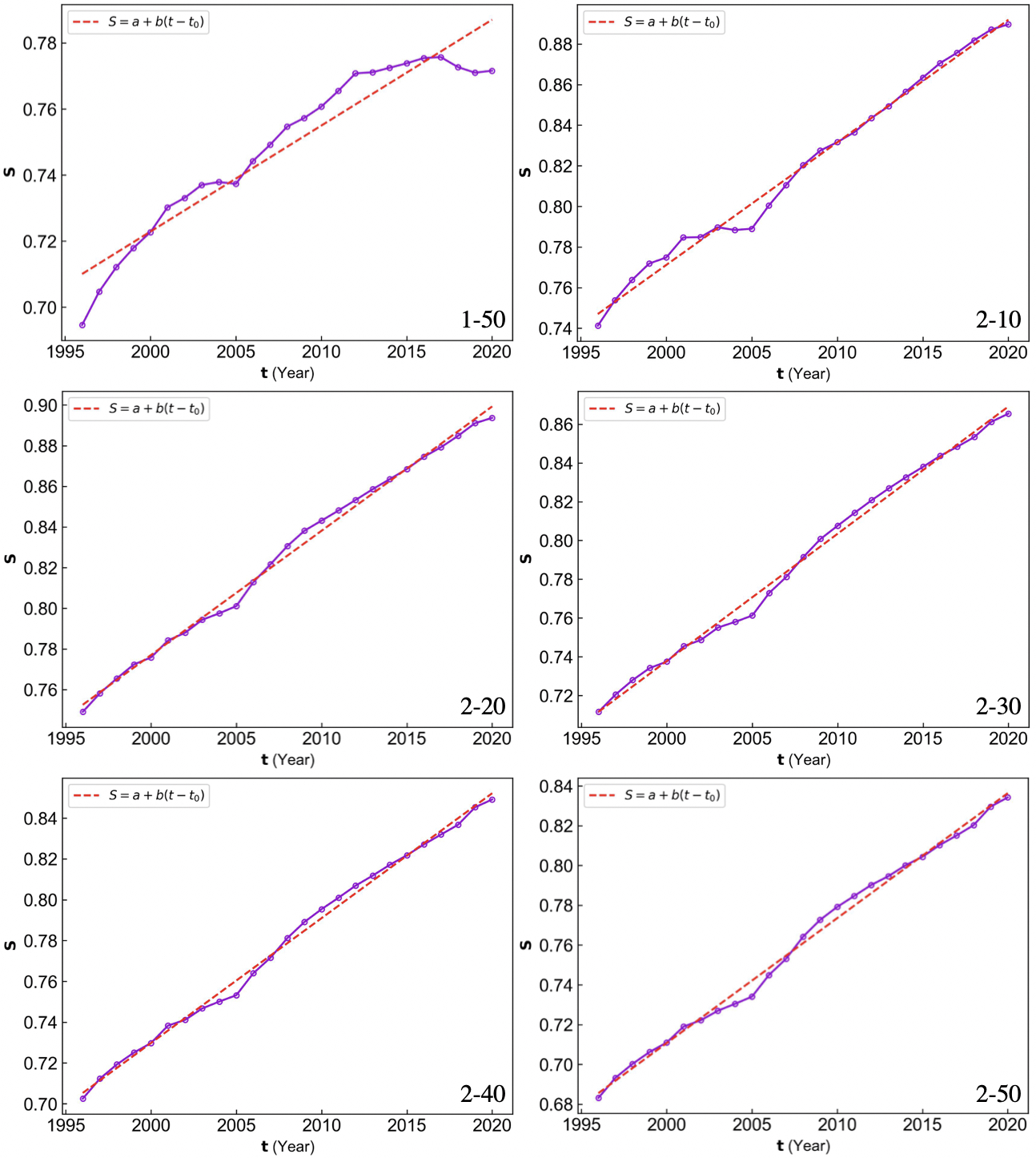}
   \caption{The variation of the entropy corresponding to the groups of countries with time. The annotated text ``1-50" in the top-left pan represents a group of countries with a range of rank from 1 (China) to 50 (Morocco). Similarly, other annotated text in each pan denotes the range of rank corresponding to the respective group of countries. The dashed lines are produced based on the results of linear regression analyses. The initial year is $t_0 =1996$.}
    \label{fig:Group_Countries}
\end{figure}
 To find out the probability distribution of annual growth rates, we have collected all the annual growth rates of all countries giving rise to the sample size of $24\times 50=1200$. The right pan of figure~\ref{fig:growth_rates} presents the histogram plots corresponding to the three types of annual growth rates. The probability distributions of annual growth rates are found to be asymmetric about the most probable values of annual growth rates giving rise to skew-symmetric distributions. The most probable values with the maximum frequency are $1.0283,0.0283$ and $0.0223$ for the relative, actual, and logarithmic annual growth rates respectively. The left part of the distribution decays to zero very fast compared to the right part which is expected as researchers in the countries are primarily eager to publish more journal articles in the next year compared to that of the current year.

\begin{table}\label{tab:CountryGroup}
\caption{The results obtained from the linear regression analyses against the time series entropy data corresponding to various groups of countries. Where ``a" and ``b" are the regression parameters. The parameter $R^2$ tests the goodness of the fitting.}
\begin{tabular}{p{0.7in} p{0.8in} p{1in} p{0.5in} p{0.4in} p{0.4in}}
Rank range & a & b & $R^2$  & $t_{disp}$ & $t_{equal}$  \\
\hline
1-50 & 0.71 $\pm$ 0.0028 & 0.0032 $\pm$ 0.0002 & 0.918 & 1775 &2086\\
2-10 & 0.747 $\pm$ 0.0017 & 0.006 $\pm$ 0.00012 & 0.99 & 1872 & 2038\\
2-20 & 0.7525 $\pm$ 0.0012 & 0.0061 $\pm$ 0.00009 & 0.995 & 1873 & 2036\\
2-30 & 0.7115 $\pm$ 0.0014 & 0.0066 $\pm$ 0.0001 & 0.994 & 1888 & 2040\\
2-40 & 0.7053 $\pm$ 0.0021 & 0.0061 $\pm$ 0.00008 & 0.996 & 1881 & 2044\\
2-50 & 0.6855 $\pm$ 0.0014 & 0.0063 $\pm$ 0.0001 & 0.994 & 1887 &2046\\
\hline
\end{tabular}
\end{table}

We have computed the entropy for countries based on their share in the total publication of the group to assess the disparity among them. Figure~\ref{fig:Group_Countries} presents the variation of the entropy corresponding to the countries in various groups as a function of time. It is found that the entropy mostly increases linearly with time implying the constant involvement of the countries in the growth of science and the increasing contribution of lagging countries. However, the top left pan of figure~\ref{fig:Group_Countries} shows that entropy continues to decay significantly after the year 2017 as the year-wise publication of China has been surging since then. And China has become a large giant in science publications. The rest of the countries lag behind it giving rise to the monopoly or disparity in the system. To investigate the equality among other countries, we therefore exclude China from other studies which are presented in the rest of the pans of figure~\ref{fig:Group_Countries}. The linear regression analyses have been performed and the parameters corresponding to these analyses are presented in table~\ref{tab:CountryGroup}. 
 The countries with ranks 2 to 10 (i.e. USA to Brazil) may contribute equally at about the year 2038 giving rise to the entropy equal to one. Similarly, $t_{equal}$ is about 2046 for all of the countries excluding China (i.e. USA to Morocco). Around this year these countries may contribute the same to the growth of science leading to spatial homogeneity.
It should be noted that for the groups of countries without China, $t_{disp}$ is in a range of years from 1872 to 1887. Around this period, a large spatial disparity was expected as few countries mostly contributed to the growth of science leading to the entropy equal to zero. Interestingly, two famous and reputed journals Nature and Science were also established in 1869 and 1880, respectively \citep{Nature_1869, Science_1880}. The establishment of two such journals is expected to give the initial momentum towards scientific research and development across the countries.

\begin{figure}
  \includegraphics[clip=true,width=\columnwidth]{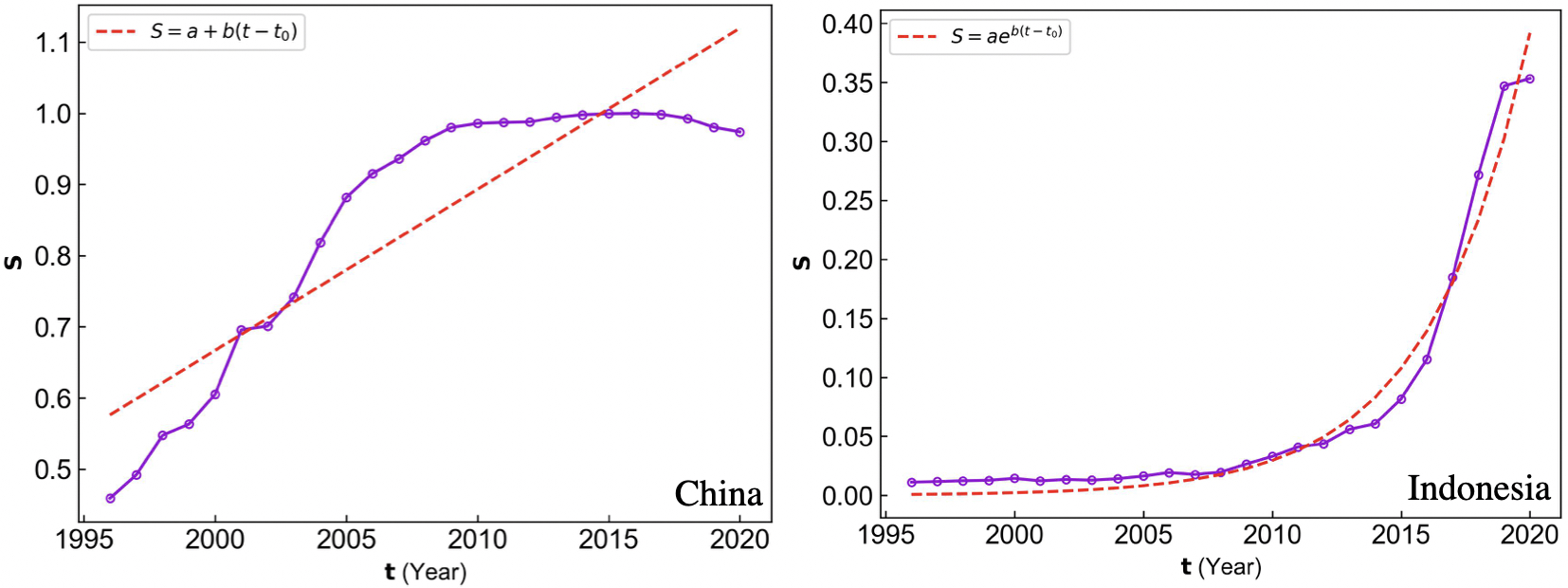}
   \caption{The time evolution of entropy between USA and other countries such as China in left pan and Indonesia in right pan. The dashed lines are produced based on the results of regression analyses. The initial year is $t_0 =1996$.}
    \label{fig:Usa_chn_idn}
\end{figure}
\begin{figure}
  \includegraphics[clip=true,width=\columnwidth]{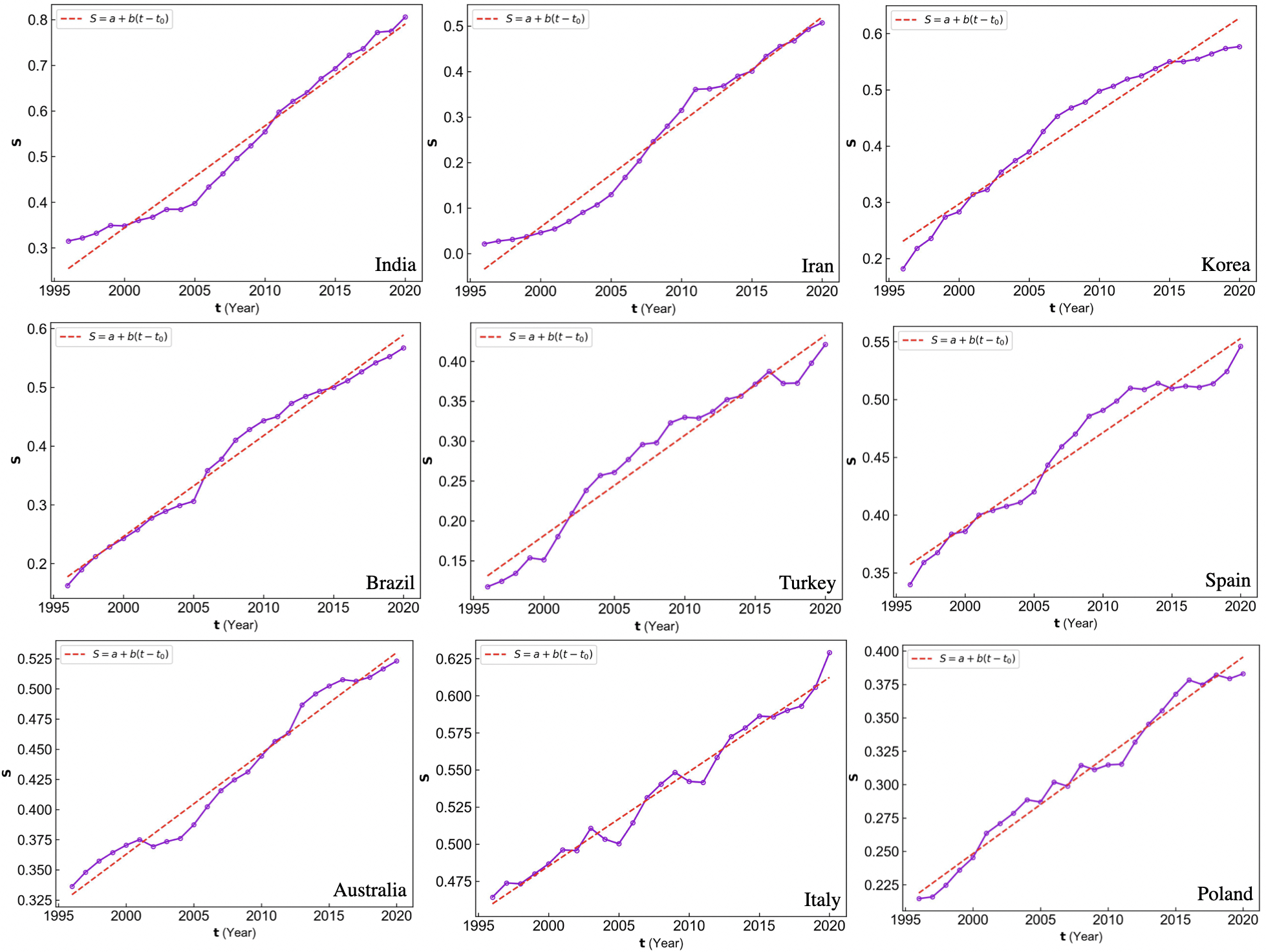}
   \caption{The variation of entropy between USA and other countries as a function of time. The dashed lines are produced based on the results of linear regression analyses. The annotated text in each pan denotes the other country that is considered for the respective study. The dashed lines are produced based on the results of linear regression analyses. The initial year is $t_0 =1996$. }
    \label{fig:Usa_linear}
\end{figure}
\begin{figure}[H]
  \includegraphics[clip=true,width=\columnwidth]{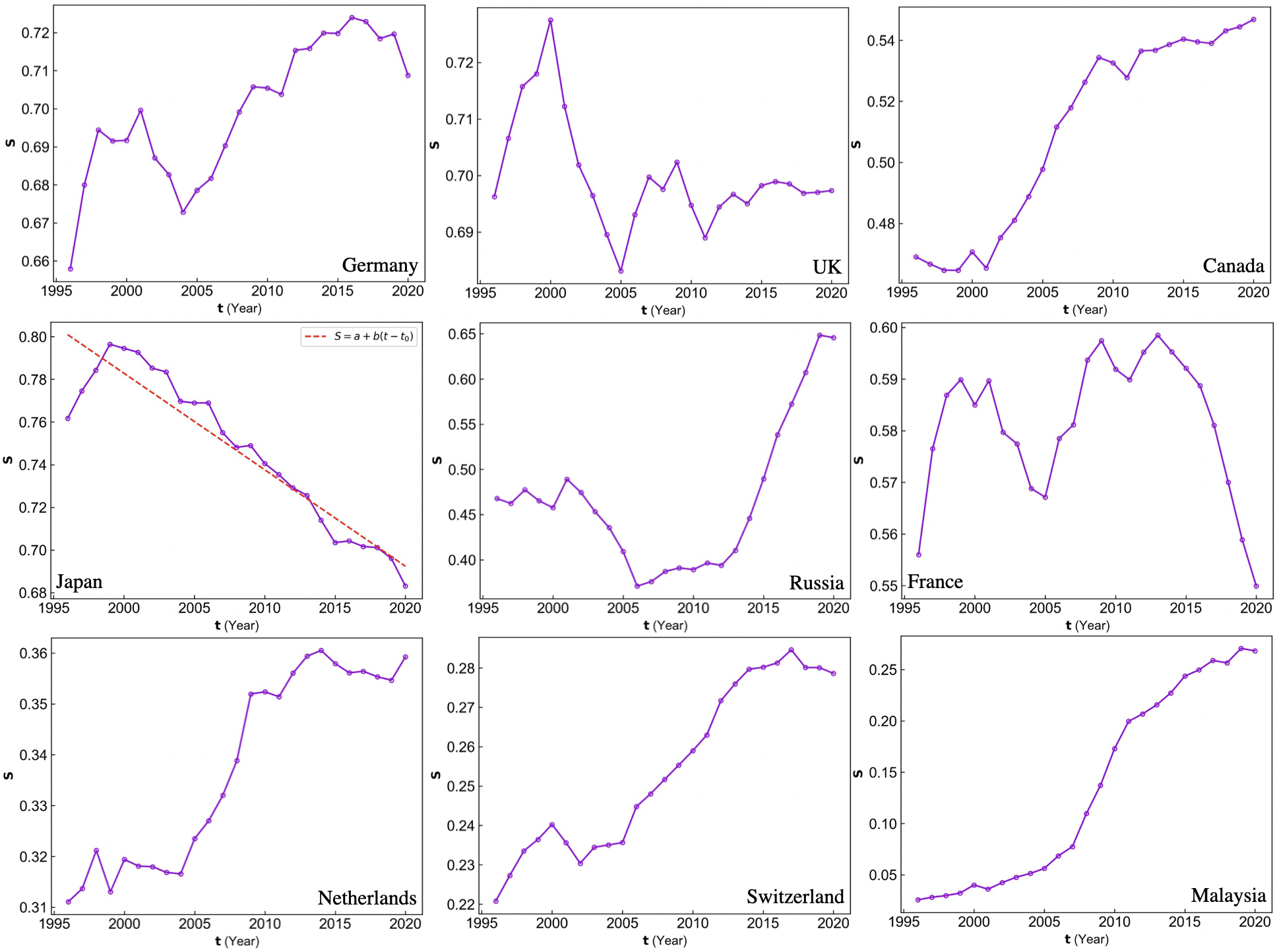}
   \caption{The variation of entropy between USA and other countries as a function of time. The dashed line is produced based on the results of linear regression analysis with the parameters $a=0.8009 \pm 0.0048$, $b=-0.0045 \pm 0.0003$, and $t_0 =1996$. }
    \label{fig:Usa_rest}
\end{figure}

It is notable that the USA was the main contributor to the growth of science up to the year 2015 and after that China holds the top position in producing science papers. To assess the stability of the current position of the USA down the road, we have computed the year-wise entropy between the USA and other countries. In these studies, we restricted the maximum rank of 20 for other countries. The left pan of figure~\ref{fig:Usa_chn_idn} presents the variation of entropy with time for the countries China and USA. Entropy increases with time initially and after attaining a maximum value equal to one at about 2015, it starts to decrease with time. This nature can be accounted for by the growth dynamics of science publications in China. The right pan of figure~\ref{fig:Usa_chn_idn} displays the evolution of entropy between the USA and Indonesia. It is found that the entropy between these two countries exponentially increases with time starting from a low value at about zero. We have estimated $t_{equal}=2024$ by utilizing the value of fitting parameters indicated in table~\ref{tab:USA}. Therefore, at the current pace, Indonesia may take a position ahead of the USA around the year 2024. 
Figure~\ref{fig:Usa_linear} shows the evolution of the entropy between the USA and some other countries with a rank less than 21. It is found that the entropy on average increases linearly with time. We have performed the linear regression analyses to estimate the time $t_{equal}$.
The parameters corresponding to the regression analysis including linear and exponential are presented in table~\ref{tab:USA}. In the year 2029, the entropy between India and the USA may equal one implying the equal contribution of these two countries in the growth of science.
After that India may hold the position ahead of the USA. Similarly, Iran may take the position ahead of the USA around the year 2041.
As can be seen in table~\ref{tab:USA}, countries except Poland are the potential candidates to earn the top rank ahead of the USA in this twenty-first century.
\begin{table}\label{tab:USA}
\caption{The results obtained from the regression analyses against the time series entropy data corresponding to the USA and other countries. Where ``a" and ``b" are the regression parameters. The parameter $R^2$ tests the goodness of the fitting. }
\begin{tabular}{p{0.7in} p{0.85in} p{1in} p{0.5in} p{0.4in} p{0.4in}}
Country & a & b & $R^2$  & $t_{disp}$ & $t_{equal}$  \\
\hline
China & 0.5761 $\pm$ 0.0331 & 0.0226 $\pm$ 0.0024 & 0.799 & 1971 & 2015\\
Indonesia & 0.0008 $\pm$ 0.0003 & 0.2588 $\pm$ 0.0157 &0.969 & 1996 & 2024\\
India & 0.2545 $\pm$ 0.0118 & 0.0223 $\pm$ 0.0008 & 0.968 & 1985 & 2029\\
Iran & -0.0342 $\pm$ 0.0107 & 0.0231 $\pm$ 0.0008 & 0.975 & 1997 & 2041\\
Korea & 0.2307 $\pm$ 0.0109 & 0.0165 $\pm$ 0.0008 & 0.951 & 1982 & 2042\\
Brazil & 0.1775 $\pm$ 0.0062 & 0.0171 $\pm$ 0.0004 & 0.985 & 1986 & 2044\\
Turkey & 0.131 $\pm$ 0.0076 & 0.0126 $\pm$ 0.0005 & 0.958 & 1986 & 2065\\
Spain & 0.3572 $\pm$ 0.0054 & 0.0081 $\pm$ 0.0004 & 0.950 & 1952 & 2075\\
Australia & 0.3295 $\pm$ 0.0041 & 0.0083 $\pm$ 0.0003 & 0.972 & 1957 & 2076\\
Italy & 0.4598 $\pm$ 0.0029 & 0.0064 $\pm$ 0.0002 & 0.976& 1924 & 2081\\
Poland & 0.2188 $\pm$ 0.0031 & 0.0074 $\pm$ 0.0002 & 0.980 & 1966 & 2102\\
\hline
\end{tabular}
\end{table}
Figure~\ref{fig:Usa_rest} shows the evolution of the entropy between the USA and the rest of the countries with a rank less than 21. It is found that the entropy varies non-linearly with time. It is therefore difficult to perform any regression analysis for determining $t_{equal}$. Also, the values of entropy in these studies are significantly low compared to the studies discussed above. Hence, it can be speculated that the position of the USA in the growth of science may remain ahead of these countries for the long term. Furthermore,  entropy corresponding to the groups USA-Japan and USA-France is found to decrease for subsequent years implying that the contribution of both countries Japan and France are not up to the mark compared to the USA. 

All of the studies presented above strongly depend on the number of year-wise scientific and technical journal articles. It should be noted that our studies do not assess the quality of science publications. The qualitative assessment is a major issue in the understanding of the actual growth of science as some researchers across the world publish substandard and fraudulent works to secure funds for research and uphold their academic position in the current ``publish or perish" environment \citep{Hanson_2023, Chapman_2019}. 
Therefore, to determine the actual trend of growth of science the quality of research and development across the countries is needed to be critically addressed. In general, the quality is often assessed by evaluating either the citation number against each published paper or the impact factor of the journals. However, both these quantities are not safeguarded as they can be manipulated \citep{Hanson_2023, Chapman_2019}.
Therefore, the investigation concerning both the quantity and quality of scientific work towards the assessment of the growth of science deserves to be future studies. It should be noted that all the prediction based on the regression analyses strongly depends on the current pace of growth of the countries. Hence, these predictions are not warranted if countries change their policy or planning in upcoming years towards scientific research and development. Nevertheless, it is found that the annual growth rate follows Taylor’s power law behavior. However, the model that can account for this observation is yet to be developed and deserves further investigation. It is therefore aimed for future study.

\section{Summary }
In summary, we have assessed the growth of science in the top fifty countries selected based on the number of year-wise scientific and technical journal articles in 2020. For the first time, we have found that three types of annual growth rates of year-wise publications exhibit a power law dependence between the mean growth rate and variance giving rise to Taylor's power law behavior. The distributions of each of these growth rates are found to follow skew-symmetric distributions. Furthermore, we have assessed the spatial disparity among the groups of countries by computing the entropy based on the contribution of each country to the total publications of the groups. For most of the studies, entropy is found to increase linearly with time giving rise to spatial homogeneity among the countries. Based on the linear regression analysis, we have found that all countries excluding China may contribute equally to the growth of science around the year 2046. In addition, by computing entropy between the USA and other countries, we have also studied the stability of the current rank of the USA against other prominent countries. Three potential countries such as Indonesia, India, and Iran may contribute much more to the growth of science than the USA around the years 2024, 2029, and 2041, respectively.

\section{Acknowledgment}
We thank Anindya Chowdhury for reading and providing critical comments to improve the manuscript.


\begin{thebibliography}{10}
\expandafter\ifx\csname url\endcsname\relax
  \def\url#1{\burl{#1}}\fi
\expandafter\ifx\csname urlprefix\endcsname\relax\def\urlprefix{URL }\fi
\providecommand{\bibinfo}[2]{#2}
\providecommand{\eprint}[2][]{\url{#2}}
\providecommand{\doi}[1]{\url{https://doi.org/#1}}
\bibcommenthead

\bibitem{Matia_2005}
\bibinfo{author}{Matia, K.}, \bibinfo{author}{Nunes~Amaral, L.~A.}, \bibinfo{author}{Luwel, M.}, \bibinfo{author}{Moed, H.~F.} \& \bibinfo{author}{Stanley, H.~E.}
\newblock \bibinfo{title}{Scaling phenomena in the growth dynamics of scientific output}.
\newblock \emph{\bibinfo{journal}{Journal of the American Society for Information Science and Technology}} \textbf{\bibinfo{volume}{56}}, \bibinfo{pages}{893--902} (\bibinfo{year}{2005}).
\newblock \urlprefix\url{https://onlinelibrary.wiley.com/doi/abs/10.1002/asi.20183}.

\bibitem{Gupta_2013}
\bibinfo{author}{Gupta, B.~M.}, \bibinfo{author}{Bala, A.} \& \bibinfo{author}{Kshitig, A.}
\newblock \bibinfo{title}{S\&t publications output of india: A scientometric analyses of publications output, 1996-2011}.
\newblock \emph{\bibinfo{journal}{Library Philosophy and Practice (e-journal)}}  (\bibinfo{year}{2013}).
\newblock \urlprefix\url{https://digitalcommons.unl.edu/libphilprac/921/}.
\newblock \bibinfo{note}{Assessed on December 2023}.

\bibitem{NSB_2020}
\bibinfo{author}{{National {S}cience {B}oard, {N}ational {S}cience {F}oundation. 2019.}}
\newblock \bibinfo{title}{{Publication Output: U.S. Trends and International Comparisons. \emph{Science and Engineering Indicators 2020}. NSB-2020-6. Alexandria, VA}}.
\newblock \urlprefix\url{https://ncses.nsf.gov/pubs/nsb20206/}.

\bibitem{Bornmann_2021}
\bibinfo{author}{Bornmann, L.}, \bibinfo{author}{Haunschild, R.} \& \bibinfo{author}{Mutz, R.}
\newblock \bibinfo{title}{Growth rates of modern science: a latent piecewise growth curve approach to model publication numbers from established and new literature databases}.
\newblock \emph{\bibinfo{journal}{Humanities and Social Sciences Communications}} \textbf{\bibinfo{volume}{8}}, \bibinfo{pages}{224} (\bibinfo{year}{2021}).
\newblock \urlprefix\url{https://doi.org/10.1057/s41599-021-00903-w}.

\bibitem{Asatani_2023}
\bibinfo{author}{Asatani, K.}, \bibinfo{author}{Oki, S.}, \bibinfo{author}{Momma, T.} \& \bibinfo{author}{Sakata, I.}
\newblock \bibinfo{title}{Quantifying progress in research topics across nations}.
\newblock \emph{\bibinfo{journal}{Scientific Reports}} \textbf{\bibinfo{volume}{13}}, \bibinfo{pages}{4759} (\bibinfo{year}{2023}).
\newblock \urlprefix\url{https://doi.org/10.1038/s41598-023-31452-8}.

\bibitem{Coccia_2018}
\bibinfo{author}{Coccia, M.}
\newblock \bibinfo{title}{General properties of the evolution of research fields: a scientometric study of human microbiome, evolutionary robotics and astrobiology}.
\newblock \emph{\bibinfo{journal}{Scientometrics}} \textbf{\bibinfo{volume}{117}}, \bibinfo{pages}{1265--1283} (\bibinfo{year}{2018}).
\newblock \urlprefix\url{https://doi.org/10.1007/s11192-018-2902-8}.

\bibitem{Patra_2006}
\bibinfo{author}{Patra, S.~K.} \& \bibinfo{author}{Mishra, S.}
\newblock \bibinfo{title}{Bibliometric study of bioinformatics literature}.
\newblock \emph{\bibinfo{journal}{Scientometrics}} \textbf{\bibinfo{volume}{67}}, \bibinfo{pages}{477--489} (\bibinfo{year}{2006}).
\newblock \urlprefix\url{https://doi.org/10.1556/Scient.67.2006.3.9}.

\bibitem{Barth_2014}
\bibinfo{author}{Barth, A.}, \bibinfo{author}{Marx, W.}, \bibinfo{author}{Bornmann, L.} \& \bibinfo{author}{Mutz, R.}
\newblock \bibinfo{title}{On the origins and the historical roots of the higgs boson research from a bibliometric perspective}.
\newblock \emph{\bibinfo{journal}{The European Physical Journal Plus}} \textbf{\bibinfo{volume}{129}}, \bibinfo{pages}{111} (\bibinfo{year}{2014}).
\newblock \urlprefix\url{https://doi.org/10.1140/epjp/i2014-14111-6}.

\bibitem{Pautasso_2012}
\bibinfo{author}{Pautasso, M.}
\newblock \bibinfo{title}{Publication growth in biological sub-fields: Patterns, predictability and sustainability}.
\newblock \emph{\bibinfo{journal}{Sustainability}} \textbf{\bibinfo{volume}{4}}, \bibinfo{pages}{3234--3247} (\bibinfo{year}{2012}).
\newblock \urlprefix\url{https://www.mdpi.com/2071-1050/4/12/3234}.

\bibitem{Bornmann_2015}
\bibinfo{author}{Bornmann, L.} \& \bibinfo{author}{Mutz, R.}
\newblock \bibinfo{title}{Growth rates of modern science: A bibliometric analysis based on the number of publications and cited references}.
\newblock \emph{\bibinfo{journal}{Journal of the Association for Information Science and Technology}} \textbf{\bibinfo{volume}{66}}, \bibinfo{pages}{2215--2222} (\bibinfo{year}{2015}).
\newblock \urlprefix\url{https://asistdl.onlinelibrary.wiley.com/doi/abs/10.1002/asi.23329}.

\bibitem{Javed_2018}
\bibinfo{author}{Javed, S.~A.} \& \bibinfo{author}{Liu, S.}
\newblock \bibinfo{title}{Predicting the research output/growth of selected countries: application of even gm (1, 1) and ndgm models}.
\newblock \emph{\bibinfo{journal}{Scientometrics}} \textbf{\bibinfo{volume}{115}}, \bibinfo{pages}{395--413} (\bibinfo{year}{2018}).
\newblock \urlprefix\url{https://doi.org/10.1007/s11192-017-2586-5}.

\bibitem{Plerou_1999}
\bibinfo{author}{Plerou, V.}, \bibinfo{author}{Amaral, L. A.~N.}, \bibinfo{author}{Gopikrishnan, P.}, \bibinfo{author}{Meyer, M.} \& \bibinfo{author}{Stanley, H.~E.}
\newblock \bibinfo{title}{Similarities between the growth dynamics of university research and of competitive economic activities}.
\newblock \emph{\bibinfo{journal}{Nature}} \textbf{\bibinfo{volume}{400}}, \bibinfo{pages}{433--437} (\bibinfo{year}{1999}).
\newblock \urlprefix\url{https://doi.org/10.1038/22719}.

\bibitem{Singh_2020}
\bibinfo{author}{Singh, P.}, \bibinfo{author}{Singh, V.~K.}, \bibinfo{author}{Arora, P.} \& \bibinfo{author}{Bhattacharya, S.}
\newblock \bibinfo{title}{India's rank and global share in scientific research -- how data sourced from different databases can produce varying outcomes} (\bibinfo{year}{2020}).
\newblock \urlprefix\url{https://arxiv.org/abs/2007.05917}.
\newblock \eprint{2007.05917}.

\bibitem{Amaral_1998}
\bibinfo{author}{Amaral, L. A.~N.}, \bibinfo{author}{Buldyrev, S.~V.}, \bibinfo{author}{Havlin, S.}, \bibinfo{author}{Salinger, M.~A.} \& \bibinfo{author}{Stanley, H.~E.}
\newblock \bibinfo{title}{Power law scaling for a system of interacting units with complex internal structure}.
\newblock \emph{\bibinfo{journal}{Phys. Rev. Lett.}} \textbf{\bibinfo{volume}{80}}, \bibinfo{pages}{1385--1388} (\bibinfo{year}{1998}).
\newblock \urlprefix\url{https://link.aps.org/doi/10.1103/PhysRevLett.80.1385}.

\bibitem{Lee_1998}
\bibinfo{author}{Lee, Y.}, \bibinfo{author}{Nunes~Amaral, L.~A.}, \bibinfo{author}{Canning, D.}, \bibinfo{author}{Meyer, M.} \& \bibinfo{author}{Stanley, H.~E.}
\newblock \bibinfo{title}{Universal features in the growth dynamics of complex organizations}.
\newblock \emph{\bibinfo{journal}{Phys. Rev. Lett.}} \textbf{\bibinfo{volume}{81}}, \bibinfo{pages}{3275--3278} (\bibinfo{year}{1998}).
\newblock \urlprefix\url{https://link.aps.org/doi/10.1103/PhysRevLett.81.3275}.

\bibitem{Gopikrishnan_1998}
\bibinfo{author}{Gopikrishnan, P.}, \bibinfo{author}{Meyer, M.}, \bibinfo{author}{Amaral, L. A.~N.} \& \bibinfo{author}{Stanley, H.~E.}
\newblock \bibinfo{title}{Inverse cubic law for the distribution of stock price variations}.
\newblock \emph{\bibinfo{journal}{The European Physical Journal B - Condensed Matter and Complex Systems}} \textbf{\bibinfo{volume}{3}}, \bibinfo{pages}{139--140} (\bibinfo{year}{1998}).
\newblock \urlprefix\url{https://doi.org/10.1007/s100510050292}.

\bibitem{Pan_2007}
\bibinfo{author}{{Pan, R. K.}} \& \bibinfo{author}{{Sinha, S.}}
\newblock \bibinfo{title}{Self-organization of price fluctuation distribution in evolving markets}.
\newblock \emph{\bibinfo{journal}{EPL}} \textbf{\bibinfo{volume}{77}}, \bibinfo{pages}{58004} (\bibinfo{year}{2007}).
\newblock \urlprefix\url{https://doi.org/10.1209/0295-5075/77/58004}.

\bibitem{Helbing_2000}
\bibinfo{author}{Helbing, D.}, \bibinfo{author}{Farkas, I.} \& \bibinfo{author}{Vicsek, T.}
\newblock \bibinfo{title}{Simulating dynamical features of escape panic}.
\newblock \emph{\bibinfo{journal}{Nature}} \textbf{\bibinfo{volume}{407}}, \bibinfo{pages}{487--490} (\bibinfo{year}{2000}).
\newblock \urlprefix\url{https://doi.org/10.1038/35035023}.

\bibitem{Silverberg_2013}
\bibinfo{author}{Silverberg, J.~L.}, \bibinfo{author}{Bierbaum, M.}, \bibinfo{author}{Sethna, J.~P.} \& \bibinfo{author}{Cohen, I.}
\newblock \bibinfo{title}{Collective motion of humans in mosh and circle pits at heavy metal concerts}.
\newblock \emph{\bibinfo{journal}{Phys. Rev. Lett.}} \textbf{\bibinfo{volume}{110}}, \bibinfo{pages}{228701} (\bibinfo{year}{2013}).
\newblock \urlprefix\url{https://link.aps.org/doi/10.1103/PhysRevLett.110.228701}.

\bibitem{Forgacs_2023}
\bibinfo{author}{Forgács, P.}, \bibinfo{author}{Libál, A.}, \bibinfo{author}{Reichhardt, C.}, \bibinfo{author}{Hengartner, N.} \& \bibinfo{author}{Reichhardt, C. J.~O.}
\newblock \bibinfo{title}{Transient pattern formation in an active matter contact poisoning model}.
\newblock \emph{\bibinfo{journal}{Communications Physics}} \textbf{\bibinfo{volume}{6}}, \bibinfo{pages}{294} (\bibinfo{year}{2023}).
\newblock \urlprefix\url{https://doi.org/10.1038/s42005-023-01387-7}.

\bibitem{Stock_2022}
\bibinfo{author}{Stock, E.~V.}, \bibinfo{author}{{da Silva}, R.} \& \bibinfo{author}{Fernandes, H.~A.}
\newblock \bibinfo{title}{A physics-based algorithm to perform predictions in football leagues}.
\newblock \emph{\bibinfo{journal}{Physica A: Statistical Mechanics and its Applications}} \textbf{\bibinfo{volume}{600}}, \bibinfo{pages}{127532} (\bibinfo{year}{2022}).
\newblock \urlprefix\url{https://www.sciencedirect.com/science/article/pii/S0378437122003740}.

\bibitem{Patra_2024_cricket}
\bibinfo{author}{Patra, D.}
\newblock \bibinfo{title}{Similarities among top one day batters: physics-based quantification} (\bibinfo{year}{2024}).
\newblock \urlprefix\url{https://arxiv.org/abs/2406.18617}.
\newblock \eprint{2406.18617}.

\bibitem{Patra_2024_meancenter}
\bibinfo{author}{Patra, D.}
\newblock \bibinfo{title}{Determination of the mean center of a region: A physics-based approach} (\bibinfo{year}{2024}).
\newblock \urlprefix\url{https://arxiv.org/abs/2406.15455}.
\newblock \eprint{2406.15455}.

\bibitem{World_Bank_2024}
\bibinfo{author}{{The World Bank}}.
\newblock \bibinfo{title}{Scientific and technical journal articles}.
\newblock \urlprefix\url{https://data.worldbank.org/indicator/IP.JRN.ARTC.SC}.
\newblock \bibinfo{note}{Data source - National Science Foundation, Science and Engineering Indicators}.

\bibitem{Zoltan_2008}
\bibinfo{author}{Zoltán~Eisler, I.~B.} \& \bibinfo{author}{Kertész, J.}
\newblock \bibinfo{title}{Fluctuation scaling in complex systems: Taylor's law and beyond1}.
\newblock \emph{\bibinfo{journal}{Advances in Physics}} \textbf{\bibinfo{volume}{57}}, \bibinfo{pages}{89--142} (\bibinfo{year}{2008}).
\newblock \urlprefix\url{https://doi.org/10.1080/00018730801893043}.

\bibitem{Vallania_2014}
\bibinfo{author}{Vallania, F. L.~M.} \emph{et~al.}
\newblock \bibinfo{title}{Origin and consequences of the relationship between protein mean and variance}.
\newblock \emph{\bibinfo{journal}{PLOS ONE}} \textbf{\bibinfo{volume}{9}}, \bibinfo{pages}{1--9} (\bibinfo{year}{2014}).
\newblock \urlprefix\url{https://doi.org/10.1371/journal.pone.0102202}.

\bibitem{Kendal_2004}
\bibinfo{author}{Kendal, W.~S.}
\newblock \bibinfo{title}{A scale invariant clustering of genes on human chromosome 7}.
\newblock \emph{\bibinfo{journal}{BMC Evolutionary Biology}} \textbf{\bibinfo{volume}{4}}, \bibinfo{pages}{3} (\bibinfo{year}{2004}).
\newblock \urlprefix\url{https://doi.org/10.1186/1471-2148-4-3}.

\bibitem{Shi_2019}
\bibinfo{author}{Shi, P.} \emph{et~al.}
\newblock \bibinfo{title}{Taylor's power law in the wenchuan earthquake sequence with fluctuation scaling}.
\newblock \emph{\bibinfo{journal}{Natural Hazards and Earth System Sciences}} \textbf{\bibinfo{volume}{19}}, \bibinfo{pages}{1119--1127} (\bibinfo{year}{2019}).
\newblock \urlprefix\url{https://nhess.copernicus.org/articles/19/1119/2019/}.

\bibitem{Giometto_2015}
\bibinfo{author}{Giometto, A.}, \bibinfo{author}{Formentin, M.}, \bibinfo{author}{Rinaldo, A.}, \bibinfo{author}{Cohen, J.~E.} \& \bibinfo{author}{Maritan, A.}
\newblock \bibinfo{title}{Sample and population exponents of generalized taylor’s law}.
\newblock \emph{\bibinfo{journal}{Proceedings of the National Academy of Sciences}} \textbf{\bibinfo{volume}{112}}, \bibinfo{pages}{7755--7760} (\bibinfo{year}{2015}).
\newblock \urlprefix\url{https://www.pnas.org/doi/abs/10.1073/pnas.1505882112}.

\bibitem{Nature_1869}
\bibinfo{title}{{History of \emph{Nature}}}.
\newblock \urlprefix\url{https://www.nature.com/nature/history-of-nature}.
\newblock \bibinfo{note}{Assessed on 5 May 2024}.

\bibitem{Science_1880}
\bibinfo{title}{{About Science \& AAAS}}.
\newblock \urlprefix\url{https://www.science.org/content/page/about-science-aaas}.
\newblock \bibinfo{note}{Assessed on 5 May 2024}.

\bibitem{Hanson_2023}
\bibinfo{author}{Hanson, M.~A.}, \bibinfo{author}{Barreiro, P.~G.}, \bibinfo{author}{Crosetto, P.} \& \bibinfo{author}{Brockington, D.}
\newblock \bibinfo{title}{The strain on scientific publishing} (\bibinfo{year}{2023}).
\newblock \urlprefix\url{https://arxiv.org/abs/2309.15884}.
\newblock \eprint{2309.15884}.

\bibitem{Chapman_2019}
\bibinfo{author}{Chapman, C.~A.} \emph{et~al.}
\newblock \bibinfo{title}{Games academics play and their consequences: how authorship, h-index and journal impact factors are shaping the future of academia}.
\newblock \emph{\bibinfo{journal}{Proceedings of the Royal Society B: Biological Sciences}} \textbf{\bibinfo{volume}{286}}, \bibinfo{pages}{20192047} (\bibinfo{year}{2019}).
\newblock \urlprefix\url{https://royalsocietypublishing.org/doi/abs/10.1098/rspb.2019.2047}.

\end{thebibliography}
\end{document}